\documentclass[epj]{svjour}
\usepackage{graphics}
\usepackage[utf8]{inputenc}  
\usepackage{graphicx,color} 
\usepackage{mathrsfs}
\usepackage{bm}
\usepackage{amsmath,amssymb,amsfonts}
\usepackage{amssymb}
\usepackage{ulem}
\usepackage{dsfont}
\usepackage{adjustbox}
\usepackage{tikz}
\usetikzlibrary{calc,trees,positioning,arrows,chains,shapes.geometric,%
    decorations.pathreplacing,decorations.pathmorphing,shapes,%
    matrix,shapes.symbols}
\usetikzlibrary{arrows,positioning} 

\tikzset{
>=stealth',
  punktchain/.style={
    rectangle, 
    rounded corners, 
    draw=black, very thick,
    text width=9em, 
    minimum height=3em, 
    text centered, 
    on chain},
  line/.style={draw, thick, <->},
  element/.style={
    tape,
    top color=white,
    bottom color=blue!50!black!60!,
    minimum width=8em,
    draw=blue!40!black!90, very thick,
    text width=10em, 
    minimum height=3.5em, 
    text centered, 
    on chain},
  every join/.style={<->, thick,shorten >=1pt},
  decoration={brace},
  tuborg/.style={decorate},
  tubnode/.style={midway, right=2pt},
}

\tikzset{
     >=stealth',
     punkt/.style={
            rectangle,
            rounded corners,
            draw=black, very thick,
            text width=6.5em,
            minimum height=2em,
            text centered},
     pil/.style={
            ->,
            thick,
            shorten <=2pt,
            shorten >=2pt,}
 }

\newcommand{\sumint}{\int\hspace{-14pt}\sum}

\begin{document}
\title{Prospective study on microscopic potential \\ with Gogny interaction}
\author{G. Blanchon\inst{1,}\thanks{e-mail: guillaume.blanchon@cea.fr} \and M. Dupuis\inst{1} \and H. F. Arellano \inst{2,1} 
 %
}                     
%
%
\institute{CEA,DAM,DIF F-91297 Arpajon, France \and Department of Physics - FCFM, University of Chile, Av. Blanco Encalada 2008, Santiago, Chile}
\date{Received: date / Revised version: date}
%
\abstract{We present our current studies and our future plans on microscopic potential based on effective nucleon-nucleon interaction and many-body 
theory. This framework treats in an unified way nuclear structure and reaction. It offers the opportunity to link the underlying effective interaction 
to nucleon scattering observables. The more consistently connected to a variety of reaction and structure experimental data the 
framework will be, the more constrained effective interaction will be. As a proof of concept, we present some recent results for both 
neutron and proton scattered from spherical target nucleus, namely $^{40}$Ca, using the Gogny D1S interaction. Possible fruitful crosstalks 
between microscopic potential, phenomenological potential and effective interaction are exposed. We then draw some prospective 
plans for the forthcoming years including scattering from spherical nuclei experiencing pairing correlations, scattering from axially 
deformed nuclei, and new effective interaction with reaction constraints. 
\PACS{
      {PACS-key}{describing text of that key}   \and
      {PACS-key}{describing text of that key}
     } 
} 
\maketitle
\section{Introduction}
\label{intro}

Producing satisfactory data evaluations based solely on many-body theories and effective nucleon-nucleon (NN) interaction is a long 
term project. Two keys to success are: \textit{(i)} robust and well tested nuclear reaction codes, such as \textsc{TALYS} 
\cite{koning_08} or \textsc{EMPIRE} \cite{herman_07}, flexible enough to incorporate new microscopic models and \textit{(ii)} 
microscopic inputs such as optical model potentials, nuclear level densities, $\gamma$-ray strength functions, and fission properties, 
all based on effective NN interaction. Advances made along this line provide more opportunities to connect the effective NN interaction to a 
broad body of structure and reaction data and as a matter of fact to improve its parametrization. 

\subsection{Nuclear energy density functional}
\begin{figure*}
\begin{center}
\adjustbox{trim={0.\width} {0.\height} {0.\width} {0.\height},clip}
{\includegraphics[width=.7\textwidth,angle=-00,clip=false]{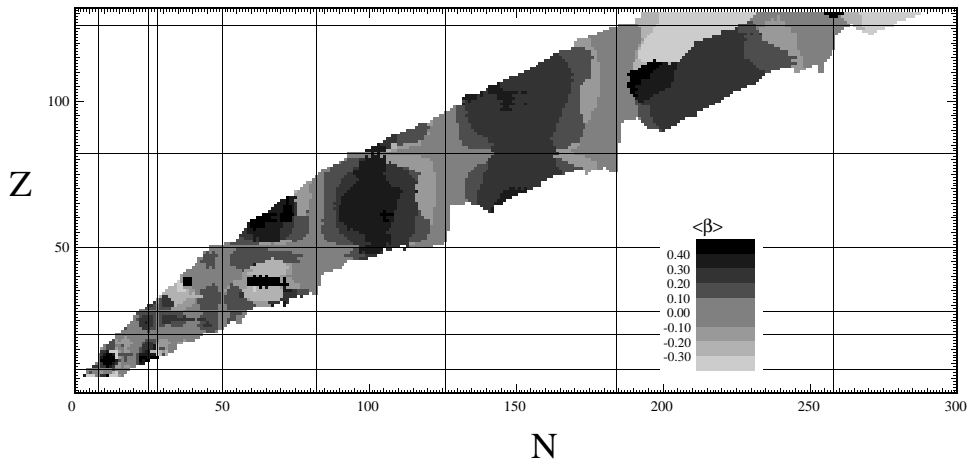}}
\caption{Chart of the nuclides showing deformations obtained with axially deformed HFB/D1S \cite{hilaire_07}.}
\label{fig:maphfb}
\end{center}
\end{figure*}

Many-body theories based on effective NN interaction such as Hartree-Fock(-Bogolyubov) (HF(B)) for static properties and (quasiparticle-)random-phase approximation 
((Q)RPA) \cite{peru_08} or five-dimension collective Hamiltonian (5DCH) \cite{libert_99} for dynamical properties 
have proven their ability to describe a wide range of nuclear structure observables, including  binding energy, charge radius, deformation, 
excitation spectrum, density and spectroscopic factors, this for nuclear masses with $A\gtrsim5$. As an illustration of the extended reach 
of the method, we show in Fig. \ref{fig:maphfb}, nuclear deformations determined within axially deformed HFB with Gogny D1S interaction 
all over the nuclide chart \cite{hilaire_07}. Effective theories are usually based on phenomenological parametrizations of the 
effective NN interaction, such as Skyrme \cite{vautherin_72,bender_03} or Gogny forces \cite{goriely_09,decharge_80,berger_91,chappert_08}. 
In the following we mainly focus on Gogny interaction. The fit of the interaction mostly relies on connection made with structure data 
through the effective theory. Those constraints are then completed by physical filters coming from infinite matter calculations. Up to now, two strategies 
have been adopted. The first and original one uses a restricted HF model where single-particle orbitals are approximated by harmonic-oscillator 
wave functions for simplicity. This makes possible to determine parameter sets of the interaction from a limited number of constraints by matrix 
inversion and from physical filters. This strategy has been applied with success to the determination of D1 \cite{decharge_80}, D1S \cite{berger_91} 
and D1N \cite{chappert_08} versions of Gogny interaction. Recent multiparticle-multihole configuration mixing studies 
\cite{pillet_12,lebloas_14} have motivated the elaboration of a generalized Gogny interaction with finite-range density, spin-orbit 
and tensor terms. Along that line, Gogny D2 interaction with a finite range density term has been designed \cite{chappert_15}. The 
second strategy is more based on brute force using HFB with a self-consistent account of quadrupole correlations energies within 
the 5DCH approach. This strategy has been used to develop the D1M parametrization of Gogny interaction which reaches a rms deviation 
with respect to the 2149 measured masses of only 798 keV \cite{goriely_09}. This improvement of the parametrization 
has been done conserving as much as possible the virtues of former Gogny interactions. 
\\
As a consequence, the global character of effective theories as well as their accuracy and their relatively low computational cost make them 
well suited to fulfill the needs for accurate nuclear data files in a reasonable time scale. 

\subsection{Nuclear codes ingredients based on effective interaction}
Various phenomenological ingredients of reaction codes have been replaced by their microscopic counter parts based on 
effective interaction. Connecting the dots, it will be possible to obtain satisfactory evaluations on the basis of reliable and 
accurate microscopic inputs only \cite{martini_14}. A good example of this on going work is the use made of results on $^{238}$U 
isotope described within QRPA with Gogny force \cite{peru_11,peru_14}. Those results have made possible a microscopic description 
of preequilibrium without \textit{ad-hoc} ``pseudo-state" prescription \cite{dupuis_12}; see also Dupuis \textit{et al.} in this issue. 
Along the same line, nuclear level densities have been obtained using a temperature-dependent HFB approach with Gogny 
interaction \cite{hilaire_12}. $\gamma-$ray strength function studies based on QRPA and Gogny interaction have also been developed \cite{martini_12,nyhus_15}. 
Concerning optical potentials, \textsc{TALYS} uses phenomenological potentials \cite{koning_03,soukhovitskii_04} in the regions where 
data are available and a modified version of the semi-microscopic Jeukenne-Lejeune-Mahaux \cite{bauge_01} elsewhere. Neither of those methods 
allows a direct connection with NN interaction. In order to fulfill this lack, we are interested in developing a microscopic 
potential based on the effective NN interaction.   

\subsection{Microscopic and ab-inito potentials} 
Depending on the projectile energy and the target mass, various strategies have been adopted in order to deal with elastic scattering 
starting from NN interaction. We now expose the pros and cons of those different methods. In the following, \textit{ab-initio} (microscopic) 
refers to methods based on bare (effective) NN interaction. \\
Nuclear matter models \cite{hufner_72} provide reasonable descriptions of nucleon elastic scattering at incident energies above about 
50~MeV \cite{dupuis_06}, even up to $\sim$1~GeV \cite{arellano_02}. The method is based on the folding of a matter density and a {g}-matrix 
effective interaction built from bare NN interaction. The density is provided by an \textit{ad-hoc} prescription such as mean-field or 
beyond mean-field approaches. Work toward a consistent treatment of both density and {g}-matrix is in progress \cite{arellano_priv}.
Recent \textit{ab-initio} calculations address the issue of reactions involving light nuclei and low-energy regime. The resonating group 
method within the no-core shell model, has successfully described nucleon scattering from light nuclei \cite{quaglioni_08}. This \textit{ab-initio} 
model has recently been extended to include three-nucleon forces for nucleon scattering from $^{4}$He \cite{hupin_13,hupin_14}. They deal 
with $^{3}$He, $^{4}$He and $^{10}$Be targets and incident energies below 15~MeV. Another method, the Green's function Monte Carlo method has been used 
to describe nucleon scattering from $^{4}$He in particular the phaseshift of the first partial waves below 5~MeV incident energy \cite{nollett_07}. 
Other \textit{ab-initio} calculations handle magic nuclei. Among them, the self-consistent Green's function (SCGF) method has been applied 
to optical potential calculations for $^{40,48,60}$Ca targets \cite{waldecker_11}. The SCGF potential is compared with a phenomenological dispersive 
potential \cite{charity_07}. This model underestimates nuclear radii and, as a consequence, is not well suited for scattering calculations. 
Further studies including three-body forces may cure this issue. Moreover work on Gorkov-Green's function theory is in progress to extend SCGF 
to nuclei around closed-shell nuclei \cite{soma_13,soma_14}. Finally, the coupled-cluster theory has been applied to proton elastic scattering from 
$^{40}$Ca \cite{hagen_12}. Cross section at 9.6~MeV and 12.44~MeV center-of-mass energy are compared with data. They observe a lack of absorption. \\
Although \textit{ab-initio} methods have made progresses in handling light and magic-nuclei, they are still yet suited neither for 
heavy targets nor for high incident energy projectiles. Another option is to build the potential starting from an effective NN interaction. 
The price to pay is to break the explicit link with bare NN interaction. The advantage is once again the extended reach of effective 
theories and the wealth of results already available. \\
The so-called nuclear structure method (NSM) for scattering \cite{vinhmau_76,bernard_79,bouyssy_81,osterfeld_81} relies on the 
self-consistent HF and RPA approximations to the microscopic optical potential \cite{vinhmau_70}. The former is a mean-field potential; the 
latter is a polarization potential built from target nucleus excitations. This method applies to double-closed shell spherical 
target nuclei well described with RPA. A strictly equivalent method, the continuum particle-vibration coupling using a Skyrme interaction, 
has been recently applied to neutron scattering from $^{16}$O \cite{mizuyama_12c}. They neglect part of the residual interaction in the coupling vertices. 
In addition, they do not address the issue of the double counting of the uncorrelated second-order diagram. Other approaches aiming at fitting a Skyrme 
effective interactions including reaction constraints are in progress, where optical potential is approximated as the HF term and the imaginary part 
of the uncorrelated particle-hole potential neglecting collectivity of target excited states \cite{pilipenko_12,xu_14}. A recent application of NSM 
with Gogny interaction is presented in Ref.~\cite{blanchon_15}. The same interaction is consistently used to generate the mean-field, the excited 
states and the couplings. In this study, special attention is given to the issue of the double counting of the uncorrelated second-order diagram. The 
subtraction of this second-order term is shown not to lead to pathological behaviors when positive incident energy is considered, contrarily to what 
is expected in Ref.~\cite{barbieri_01}. Moreover, the use of the finite-range Gogny interaction prevents from the necessity of \textit{ad-hoc} momentum 
cut-off when second-order effects are considered. 

\subsection{Project}

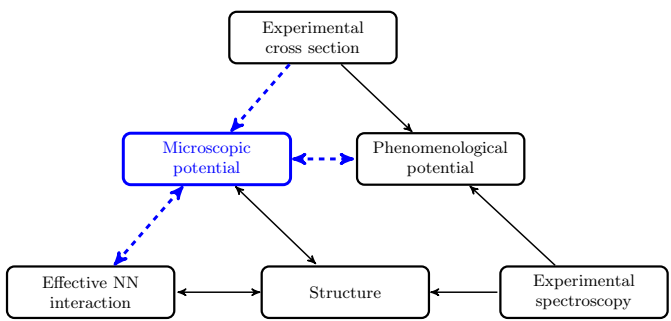
\begin{figure}[h!]
\centering
\scalebox{0.7}{
\begin{tikzpicture}
  [node distance=.8cm,
  start chain=going below,]
     \node[punktchain,join,] (exp) {Experimental cross section};
     \node (micro) [punktchain, on chain=going left, yshift = -2.3cm,xshift = 2.cm,blue,ultra thick] {Microscopic\\  potential};
     \draw[->, dashed,blue,ultra thick] (exp) -- (micro);
      \begin{scope}[start branch=hoejre,]
      \node (pheno) [punktchain, on chain=going right, xshift = .4cm,join=by {<->, dashed,blue,ultra thick},] {Phenomenological potential};
    \end{scope}
     \draw[->, thick,] (exp) -- (pheno);
      \node (struct) [punktchain, join, yshift = -.7cm ,xshift = 2.6cm]  {Structure};
      \begin{scope}[start branch=hoejre,]
      \node (spectro) [punktchain, on chain=going right, yshift = -0.cm ,xshift = .5cm,join=by {<-}] {Experimental spectroscopy};
    \end{scope}
        \draw[->, thick,] (spectro) -- (pheno);
    \begin{scope}[start branch=hoejre,]
      \node (spectro) [punktchain, on chain=going left, yshift = -0.cm , xshift = -.8cm,join=by {<->}] {Effective NN interaction};
    \end{scope}
    \draw[<->, dashed,ultra thick,blue] (spectro) -- (micro);
  \end{tikzpicture}}
\caption{Connections. Dashed arrows emphasize connections discussed in the following. }
\label{fig:connect}
\end{figure}

In the continuity of work presented in Ref.~\cite{blanchon_15}, we wish to explore the possible connections provided by a microscopic optical potential using NSM. 
The framework is summarized in Fig.~\ref{fig:connect}. 
In the first line of the diagram, effective interaction is used within structure models to make spectroscopic predictions. Feedback from 
experiment provides constraints on the interaction whenever a reliable structure model is used. Effective interaction and structure calculation 
can then be used to define the microscopic potential through NSM. Once the corresponding scattering problem is solved, feedbacks are made 
possible from cross section data providing reaction constraints to the fit of the interaction. Microscopic potentials are nonlocal, complex 
and energy dependent. They can provide prescriptions for future phenomenological potentials, in particular concerning the shape of the 
nonlocality, the energy dependence. One can as well investigate the origin of the volume and the surface part of the potential in terms of 
target excitations. Reciprocally, when data are available, phenomenological potentials can help identifying contributions missing in 
microscopic potentials. Moreover whenever phenomenological potentials obey dispersion relation, a connection is also made with spectroscopy. 
In Fig.~\ref{fig:connect}, the main connections we wish to investigate in the following are highlighted. \\

In the following, the NSM formalism for spherical-target nuclei and the integro-differential Schrödinger equation are briefly exposed in Sec.~\ref{sec:formalism} 
and Sec.~\ref{sec:schro}, respectively. In Sec.~\ref{sec:reso}, we emphasize the importance of the exact treatment of the intermediate HF propagator 
and more precisely the account of single-particle resonances. As a proof of concept in Sec.~\ref{sec:cross}, we apply NSM to nucleon elastic 
scattering from $^{40}$Ca. Some possible crosstalks between phenomenological potentials and their microscopic counter part are 
discussed in Sec.~\ref{sec:pot}. In Sec.~\ref{sec:int}, we show how phenomenological nonlocal potential can relate to the effective NN 
interaction through volume integrals. Finally in Sec.~\ref{sec:next}, we draw plans for the decade to come. In particular, 
we mention the issue of spherical target nuclei experiencing pairing correlations and the one of deformed target nuclei.

\section{Method for spherical target}
\label{sec:method}

\subsection{NSM potential}
\label{sec:formalism}
\begin{figure}[h!]
\centering
\begin{tikzpicture}[node distance=1cm, auto,]
  \node[] (market) {Schrödinger equation};
 \node[above=of market] (dummy) {};
 \node[right=of dummy,xshift=-0.5cm] (t) {$V^{HF}(\rho)$};
 \node[left=of dummy,xshift=-0.5cm] (g) {$\rho$}
   edge[pil,->, bend left=45] node[auto] {{\small NN interaction}} (t)
   edge[pil,<-, bend right=45] node[auto] {} (t);
\end{tikzpicture}
\caption{Self-consistent Hartree-Fock.}
\label{fig:iter-hf}
\end{figure}
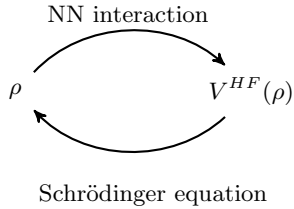

The NSM formalism is presented in detail in Ref.~\cite{vinhmau_70}. We briefly introduce the key points of the formalism. Equations 
are presented omitting spin for simplicity. The potential, $V$, consists 
of two components,
\begin{equation}
 V = V^{HF} + \Delta V. \label{eq:v}
\end{equation}
This potential will be referred to as the NSM potential in the following. The former is a mean-field potential; the 
latter is a polarization potential built from target nucleus excitations. The HF potential, $V^{HF}$, is the major 
contribution to the real part of the optical potential. The polarization potential, $\Delta V$, brings 
only a correction to the real part of $V$ and entirely generates its imaginary contribution. The HF potential is 
obtained in the Green's function formalism neglecting 
two-body correlations \cite{kadanoff_62}. It reads, 
\begin{eqnarray}
V^{HF}(\textbf{r},\textbf{r'})= \int d\textbf{r}_{1} v(\textbf{r},\textbf{r}_{1}) \rho(\textbf{r}_{1})\delta(\textbf{r}-\textbf{r'})-v(\textbf{r},\textbf{r'})\rho(\textbf{r},\textbf{r'}),\nonumber \\
\label{eq:vhf}
\end{eqnarray}
where $v$ is the effective NN interaction and 
\begin{eqnarray}
 \rho(\textbf{r}) &=& \sum_{i} |\phi_{i}(\textbf{r})|^{2}, \\
 \rho(\textbf{r},\textbf{r'}) &=& \sum_{i} \phi_{i}^{*}(\textbf{r}) \phi_{i}(\textbf{r'}), 
\label{eq:denshf}
\end{eqnarray}
are the local and nonlocal densities with $i$ running over occupied states. $V^{HF}$ is made of a local direct term 
and an exchange term which is nonlocal because of the finite range of Gogny interaction. It is energy independent. Rearrangement 
contributions stemming from the density-dependent term of the interaction are also taken into account. Together with 
Schrödinger equation, Eq.~\eqref{eq:vhf} defines a self-consistent scheme as shown in Fig.~\ref{fig:iter-hf}.

\begin{figure}[h!]
\centering
\begin{tikzpicture}[node distance=1cm, auto,]
  \node[] (market) {Schrödinger equation};
 \node[above=of market] (dummy) {};
 \node[right=of dummy,xshift=-1.cm] (t) {$V^{HF}(\rho)+\Delta V(\rho)$};
 \node[left=of dummy,xshift=-1.cm] (g) {$\rho$}
   edge[pil,dashed,->, bend left=45] node[auto] {{\small NN interaction}} (t)
   edge[pil,<-, bend right=45] node[auto] {} (t);
\end{tikzpicture}
\caption{Self-consistent RPA (full-line and dashed-lined arrows). Consistent RPA on top of a self-consistent HF (full-line arrow only). 
The HF propagator is dressed only once.}
\label{fig:iter-rpa}
\end{figure}
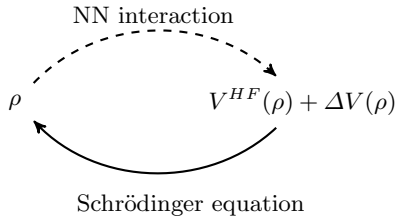

\begin{figure}[h!]
\centering
\begin{minipage}[c]{0.4\linewidth}
\centering
\adjustbox{trim={0.\width} {0.\height} {0.\width} {0.\height},clip}
{\includegraphics[width=.9\textwidth,angle=-00,clip=false]{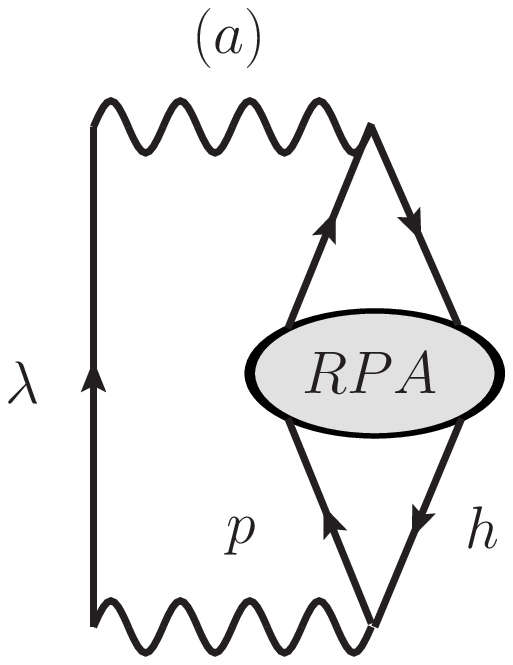}}
\adjustbox{trim={0.\width} {0.\height} {0.\width} {0.\height},clip}
{\includegraphics[width=1.\textwidth,angle=-00,clip=false]{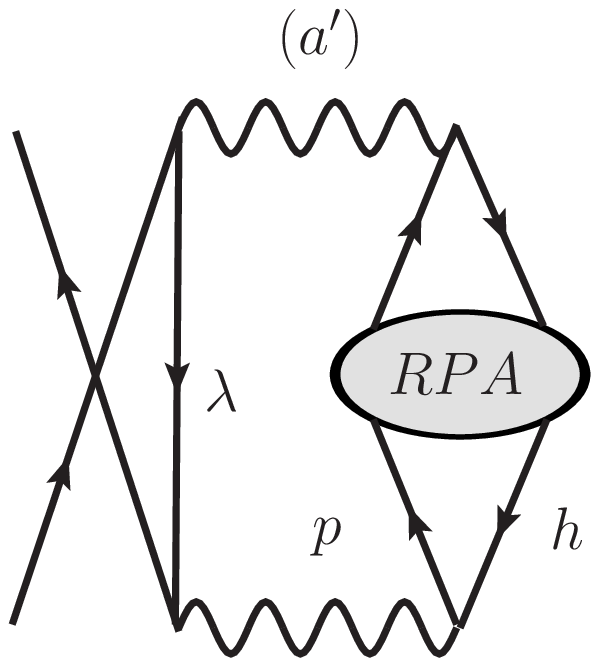}}
\end{minipage}
\begin{minipage}[c]{0.45\linewidth}
\centering
\adjustbox{trim={0.\width} {0.\height} {0.\width} {0.\height},clip}
{\includegraphics[width=.9\textwidth,angle=-00,clip=false]{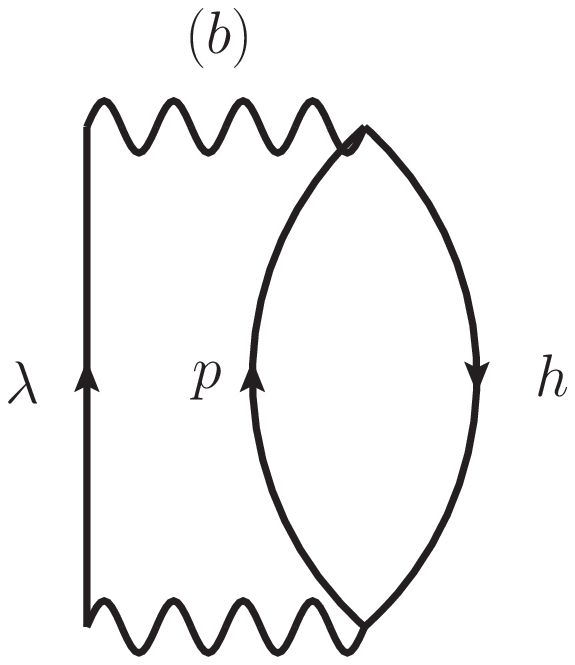}}
\adjustbox{trim={0.\width} {0.\height} {0.\width} {0.\height},clip}
{\includegraphics[width=1.\textwidth,angle=-00,clip=false]{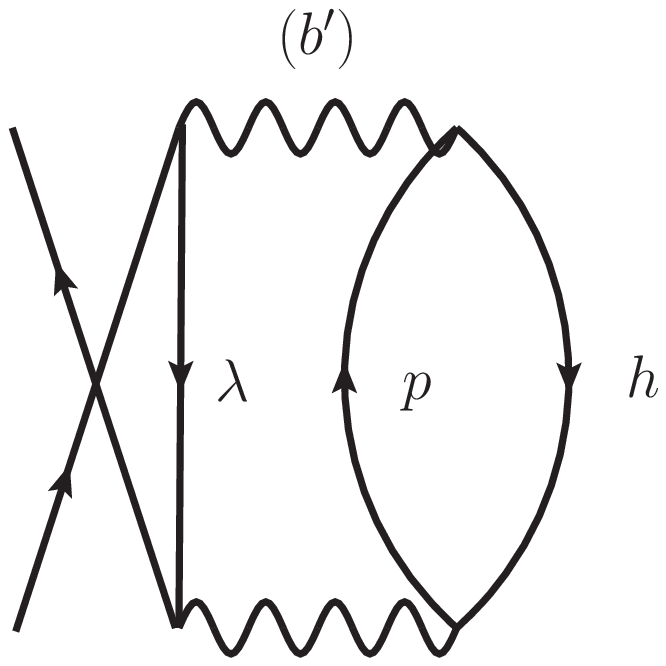}}
\end{minipage}
\caption{Diagrammatic contributions of $V^{RPA}$ (a) and $V^{(2)}$ (b). Indices $p$, $h$ and $\lambda$ refer to 
particle, hole and the intermediate state in the HF field, respectively. Wavy lines stand for the effective NN interaction.}
\label{fig:diag}
\end{figure}

The polarization potential, $\Delta V$ in Eq.~\eqref{eq:v}, is built coupling the elastic channel to the intermediate excited states of the 
target nucleus. Those excited states are described within the RPA formalism. Both excited states and couplings are generated 
using Gogny interaction. Together with Schrödinger equation, Eq.~\eqref{eq:v} now defines a new self-consistent scheme, illustrated 
in Fig.~\ref{fig:iter-rpa} when considering both full-line and dashed-line arrows.
In practice, as described in Fig.~\ref{fig:iter-rpa}, we first converge the HF scheme, as shown in Fig.~\ref{fig:iter-hf}, then we 
dress only once the HF propagator by coupling to the excitations of the target. This makes the scheme only consistent at that stage. Going 
into more details, the polarization contribution to the potential reads,
\begin{equation}
 \Delta V = V^{PP}+V^{RPA}-2V^{(2)}, \label{eq:dv}
\end{equation}
where $V^{PP}$ and $V^{RPA}$ are contributions from particle-particle and particle-hole correlations, respectively. The 
uncorrelated particle-hole contribution, $V^{(2)}$, is accounted for once in $V^{PP}$ and twice in $V^{RPA}$. When 
two-body correlations are neglected in Eq.~\eqref{eq:dv}, one expects $\Delta V$ to reduce to $V^{(2)}$. As a matter of 
fact $V^{(2)}$ shall be subtracted twice \cite{vinhmau_70}. This formalism takes into account all the correlations explicitly. 
Although it is well suited for \textit{ab-initio} developments, we wish to make the connection with effective NN interactions. 
In practice, if one uses an effective interaction with a density-dependent term, such as Gogny or Skyrme forces, attention 
must be paid to correlations already accounted for in the interaction \cite{bouyssy_81}. Indeed, in such a case, part of particle-particle 
correlations is already contained at the HF level. We thus use the same prescription as in Ref.~\cite{bernard_79}, 
omitting the real part of $V^{PP}$ while approximating the imaginary part of $V^{PP}$ by $\textrm{Im}\left[V^{(2)}\right]$. 
Then Eq.~\eqref{eq:dv} reduces to
\begin{equation}
 \Delta V = \textrm{Im}\left[V^{(2)}\right]+V^{RPA}-2V^{(2)}. \label{eq:sig-approx}
\end{equation}
Both ingredients of Eq.~\eqref{eq:sig-approx}, $V^{RPA}$ and $V^{(2)}$, can be expressed in terms of diagrams. In Fig.~\ref{fig:diag}, wavy lines stand 
for the effective NN interaction and up (down) arrows stand for HF particle (hole) propagators. Subscript $p$ ($h$) refers 
to the quantum numbers of the single-particle (hole) HF states used to build target excitations. The subscript $\lambda$ refers to 
the quantum numbers of the intermediate single-particle state of the scattered nucleon. Both discrete and continuum spectra 
of the intermediate single-particle state are accounted for. Label (a) refers to $V^{RPA}$ 
built with an unoccupied intermediate state. Label (a') refers to $V^{RPA}$ built with an occupied intermediate state. 
Labels (b) and (b') refer to the corresponding uncorrelated particle-hole contributions. Exchange diagrams are taken into 
account. \\

\noindent For nucleons with incident energy $E$, the RPA potential, corresponding to Figs.~\ref{fig:diag}a and \ref{fig:diag}a', reads,
\begin{eqnarray}
  V^{RPA}({\bf r,r'},E)&=& \sum_{N\neq 0} \sumint_{\lambda}
  \bigg[{{n_{\lambda}}\over{E-\varepsilon_{\lambda}+E_{N}-i\Gamma(E_{N})}} \nonumber\\
  &+&{{1-n_{\lambda}}\over{E-\varepsilon_{\lambda}-E_{N}+i\Gamma(E_{N})}}\bigg]\nonumber\\
  &\times& \Omega^{N}_{\lambda}(\textbf{r})\Omega^{N}_{\lambda}(\textbf{r'}), \label{eq:vrpa}
\end{eqnarray}
where $n_{i}$ and $\varepsilon_{i}$ are occupation number and energy of the single-particle state $\phi_{i}$ in the HF 
field, respectively. $E_{N}$ and $\Gamma(E_{N})$ represent the energy and the width of the $N^{th}$ excited state of the 
target, respectively. Additionally, 
\begin{eqnarray}
 \Omega^{N}_{\lambda}(\textbf{r}) = \sum_{(p,h)} \left[X^{N,(p,h)}F_{ph\lambda}(\textbf{r}) + Y^{N,(p,h)}F_{hp\lambda}(\textbf{r})\right],
\end{eqnarray}
where $X$ and $Y$ denote the usual RPA amplitudes and
\begin{equation}
 F_{ij\lambda}(\textbf{r}) = \int d^{3} \textbf{r}_{1} \phi^{*}_{i}(\textbf{r}_{1})v(\textbf{r},\textbf{r}_{1})\left[1-\textrm{\^P}\right]\phi_{\lambda}(\textbf{r})\phi_{j}(\textbf{r}_{1}),
 \label{eq:fij}
\end{equation}
where $\textrm{\^P}$ is a particle-exchange operator and $v$ is the same effective NN interaction as in Eq.~\eqref{eq:vhf}. The uncorrelated particle-hole 
contribution, corresponding to Figs.~\ref{fig:diag}b and \ref{fig:diag}b', reads
\begin{eqnarray}
  V^{(2)}({\bf r,r'},E) &=& \frac{1}{2}\sum_{ij} \sumint_{\lambda} \bigg[{{n_{i}(1-n_{j})n_{\lambda}}\over{E-\varepsilon_{\lambda}+E_{ij}-i\Gamma(E_{ij})}}\nonumber\\
  &+&{{n_{j}(1-n_{i})(1-n_{\lambda})}\over{E-\varepsilon_{\lambda}-E_{ij}+i\Gamma(E_{ij})}}\bigg]\nonumber\\
  &\times&  F_{ij\lambda}(\textbf{r})F^{*}_{ij\lambda}(\textbf{r}'), \label{eq:vph}
\end{eqnarray}
with $E_{ij} = \varepsilon_{i}-\varepsilon_{j}$, the uncorrelated particle-hole energy. \\

In practice, $V^{HF}$ is determined in coordinate space to ensure the correct asymptotic behavior of single-particle states. This 
nonlocal potential is then used to build the single-particle intermediate state used to determine the polarization potential, 
$\Delta V$. It is worth mentioning that the HF potential in coordinate space reproduces bound state energies obtained with the HF/D1S code on oscillator basis 
which makes our result reliable.\\
The description of target excitations is obtained solving RPA equations in a harmonic oscillator basis, including fifteen major shells 
\cite{blaizot_77} and using the Gogny D1S interaction \cite{berger_91}. We account for RPA excited states with spin up to $J=8$, including 
both parities, in order to achieve the convergence of cross section calculations. The first $J^{\pi}=1^{-}$ excited state given by RPA, containing 
the spurious translational mode, is removed from the calculation. Moreover, in order to avoid spurious modes in the uncorrelated particle-hole 
term, we approximate the $J^{\pi}=1^{-}$ contribution in $V^{(2)}$ by half that of the $J^{\pi}=1^{-}$ contribution in $V^{RPA}$. \\
Even though RPA/D1S method provides a good overall description of the spectroscopic properties of double-closed shell nuclei, still some contributions 
are left out. First, the projection on an oscillator basis discretizes the RPA continuum. As a consequence,	 the escape width is missing from 
the structure calculation. Second, couplings to two or more particle-hole states are excluded from the model space even though they may 
play a significant role. The impact of these couplings is a strength redistribution through a damping width as well as a shift in energy 
of excited states. Third, the optical potential is, by definition, built to provide the energy-averaged $S$-matrix. Hence, the rapid 
fluctuations, the potential exhibits at low energy due to compound-elastic contribution, shall be averaged before identifying 
the result of Eq.\eqref{eq:v} with an optical potential \cite{feshbach_58}. \\
In the present work, we simulate those three different widths assigning a single phenomenological width, $\Gamma(E_{N})$, to each RPA state. 
$\Gamma(E_{N})$ takes the value of 2, 5, 15 and 50~MeV, for excitation energies of 20, 50, 100 and 200~MeV, respectively. Those values 
have not been fitted in order to better reproduce cross section data. Longer term solutions are planned in order to provide more microscopic 
prescriptions for those widths. The escape width can be obtained using continuum RPA \cite{dedonno_11,mizuyama_12c}. We also plan to 
determine the damping width and the energy shift using the multiparticle-multihole configuration mixing method \cite{pillet_08}.

\subsection{Integro-differential Schr\"odinger equation}
\label{sec:schro}
The integro-differential Schr\"odinger equation for bound states and scattering is solved without localization procedures. The radial Schr\"odinger equation reads,
\begin{eqnarray}
&&\hspace{-1.cm}-{{\hbar^{2}}\over{2\mu}} \left[ \dfrac{d^{2}}{dr^{2}}-\frac{l(l+1)}{r^2}\right]f_{lj}(r) \nonumber \\ 
 &&\hspace{.7cm}+ r \int \nu_{lj}(r,r';E) f_{lj}(r') r'dr'= E f_{lj}(r), \label{eq:schro}
\end{eqnarray}
where $f_{lj}(r) = r \phi_{lj}(r)$ is the partial wave for the projectile-target relative motion, $E$ is the 
incident nucleon energy and $\nu_{lj}(r,r';E)$ is defined from the multipole expansion of the nonlocal potential
\begin{equation}
V(\textbf{r}\sigma,\textbf{r'}\sigma';E) = \sum_{ljm} {\cal Y}_{ljm}({\bf \hat r} \sigma) \nu_{lj}(r,r';E){\cal Y}_{ljm}^{\dagger}({\bf \hat r'} \sigma'), \label{eq:pw_exp}
\end{equation}
with
\begin{equation}
{\cal Y}_{ljm}~({\bf \hat r} \sigma)~\equiv~[Y_{l}({\bf \hat r}) \otimes \chi_{1/2}(\sigma)]_{jm}.
\end{equation}
The potential, $V$, is complex and energy dependent for $E>0$, and real and energy independent for $E<0$. Discrete solutions are obtained 
by expressing Eq.~\eqref{eq:schro} on a mesh in coordinate space and performing the corresponding matrix diagonalization \cite{hooverman_72}. 
For positive energies, the scattering problem with the correct asymptotic conditions turns into a matrix inversion following J. Raynal's method 
for scattering exposed in the \textsc{DWBA} code explanatory leaflet \cite{raynal_98}. 

\subsection{Resonances in the intermediate wave}
\label{sec:reso}
\begin{figure}[h!]
\centering
\begin{minipage}[c]{0.49\linewidth}
\adjustbox{trim={0.\width} {0.\height} {0.\width} {0.\height},clip}
{\includegraphics[width=1.\textwidth,angle=-90,clip=false]{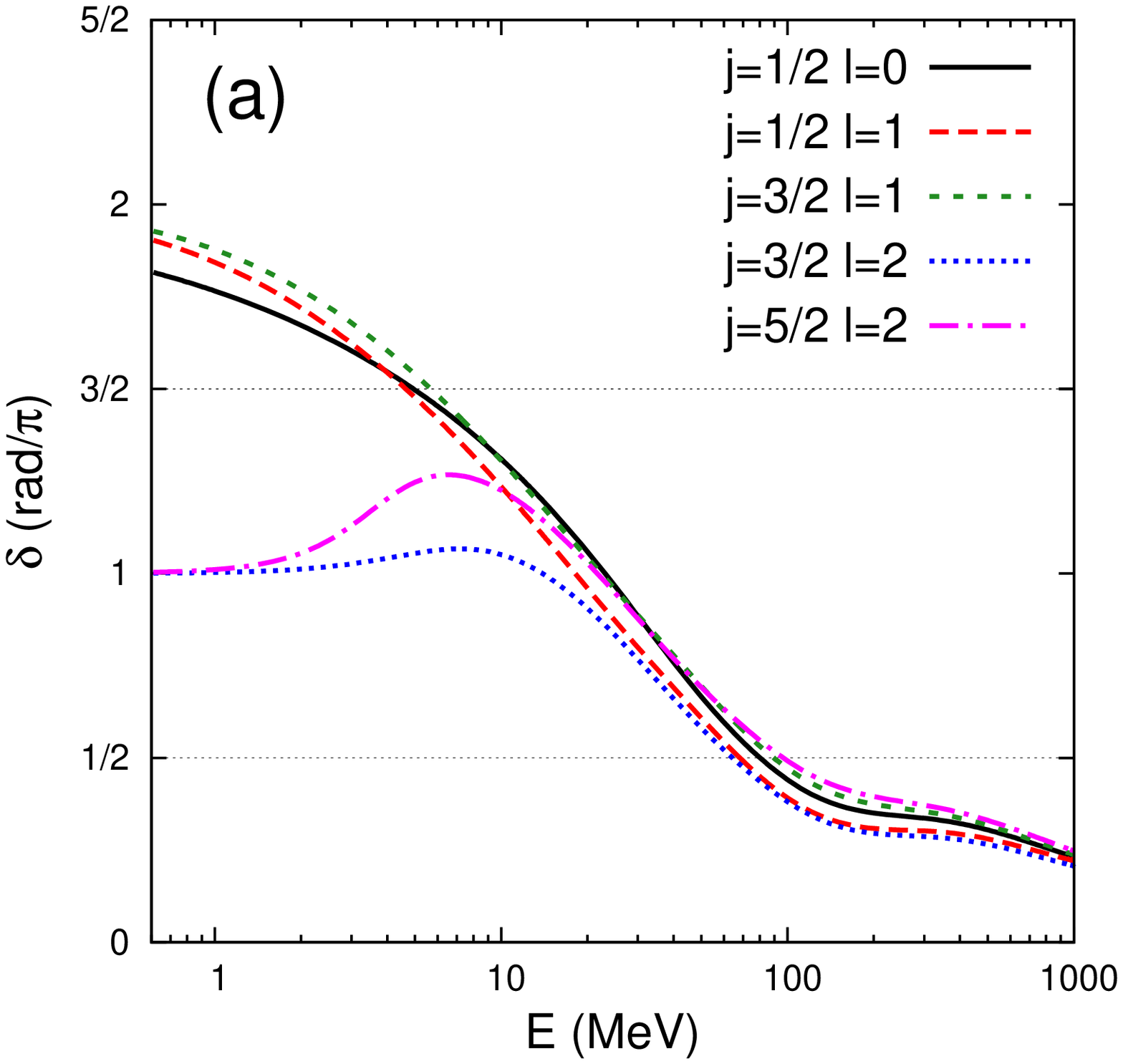}}
\adjustbox{trim={0.\width} {0.\height} {0.\width} {0.\height},clip}
{\includegraphics[width=1.\textwidth,angle=-90,clip=false]{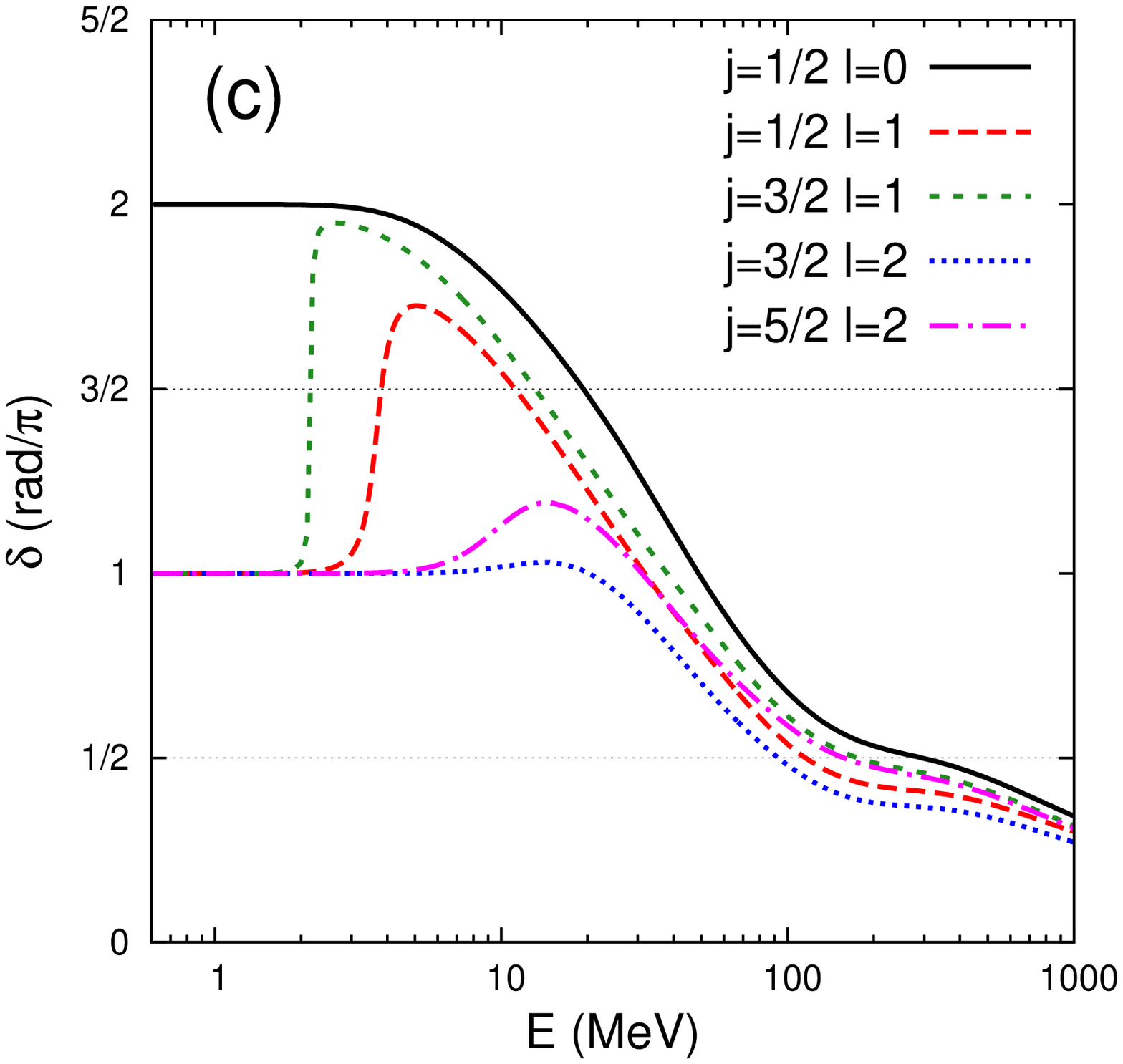}}
\end{minipage}
\hfill
\begin{minipage}[c]{0.49\linewidth}
\adjustbox{trim={0.\width} {0.\height} {0.\width} {0.\height},clip}
{\includegraphics[width=1.\textwidth,angle=-90,clip=false]{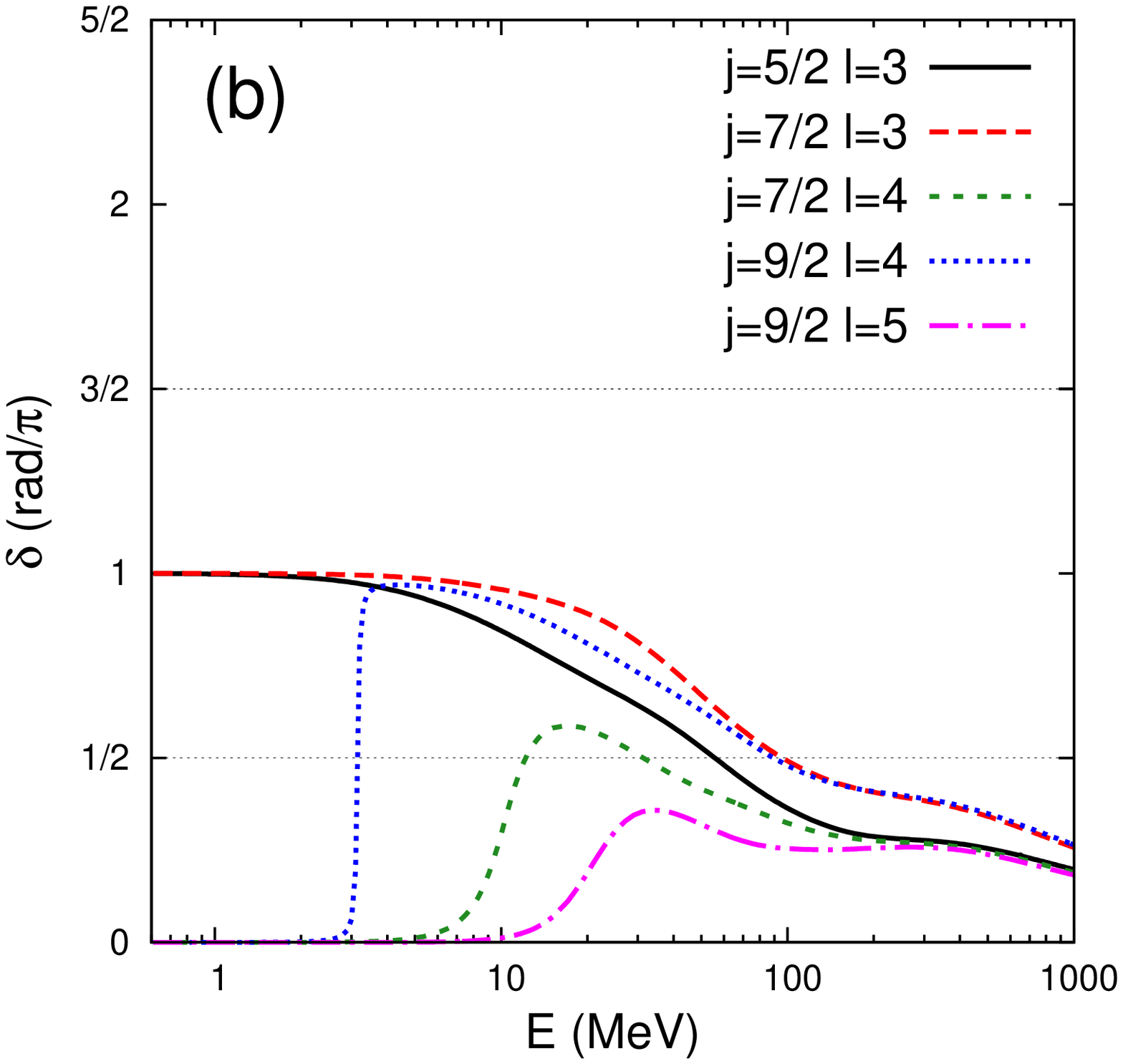}}
\adjustbox{trim={0.\width} {0.\height} {0.\width} {0.\height},clip}
{\includegraphics[width=1.\textwidth,angle=-90,clip=false]{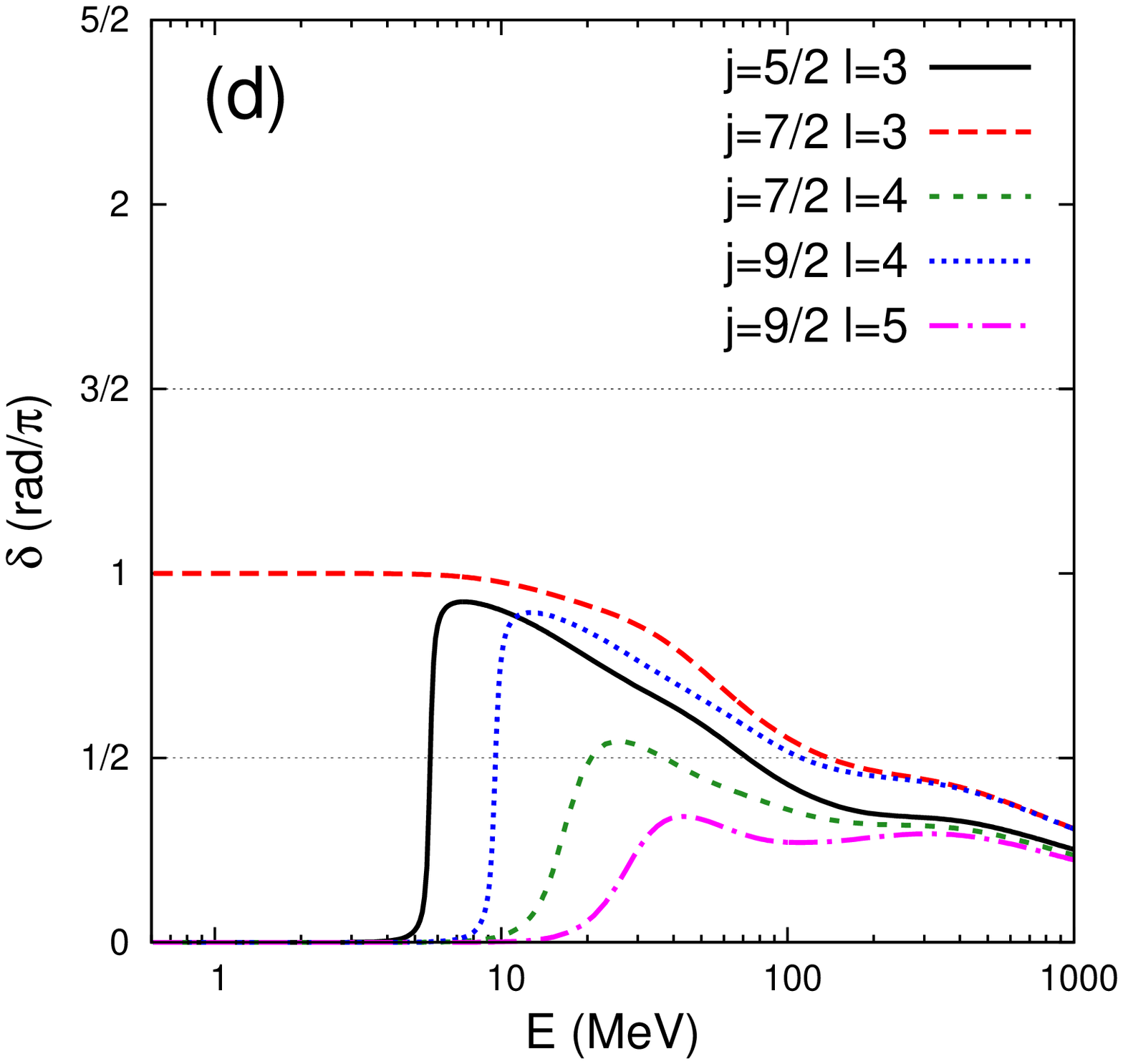}}
\end{minipage}
\caption{Phaseshift in the HF field for neutron (panel (a) and (b)) and proton (panel (c) and (d)) scattering from $^{40}$Ca as a 
function of incident energy for the first ten partial waves.}
\label{fig:phasehf}
\end{figure}

We now would like to emphasize the impact of an exact treatment of $\phi_{\lambda}$ on the second-order terms of the potential, $V^{RPA}$ and $V^{(2)}$ 
(see Eqs.~\eqref{eq:vrpa} through \eqref{eq:vph}); especially, the role of single-particle resonances already discussed by Rao \textit{et al.} \cite{rao_73} 
within a phenomenological approach. In previous works, $\phi_{\lambda}$ has often been approximated by a plane wave for neutron and a Coulomb wave for proton 
\cite{bouyssy_81} or discretized \cite{bernard_79}. In this work, we include both discrete and continuum spectra of $\phi_{\lambda}$ determined in the HF 
field with the correct asymptotic solutions. Phaseshifts for neutron and proton scattering from $^{40}$Ca in the HF field are shown in Fig.~\ref{fig:phasehf}. 
We observe single-particle resonances for several partial waves each time phaseshift increases rapidly through an odd multiple of $\pi/2$. Resonance energies 
are summarized in Table~\ref{table:res}.
\begin{table}[h!]
\caption {HF single-particle resonance energies (in MeV) for neutron (n) and proton (p) scattering from $^{40}$Ca.}
\begin{center}
\begin{tabular}{|c|c|}
\hline
          n                      &      p                            \\
\hline
  12.18   \textit{(j=7/2, l=4) } &    3.70    \textit{(j=1/2,  l=1)} \\
   3.15   \textit{(j=9/2, l=4)} &    2.15    \textit{(j=3/1,  l=1)} \\
  14.87   \textit{(j=11/2, l=5)} &    5.65    \textit{(j=5/2,  l=3)} \\
  31.09   \textit{(j=13/2, l=6)} &   20.69    \textit{(j=7/2,  l=4)} \\
                                 &    9.55    \textit{(j=9/2,  l=4)} \\
                                 &   22.08    \textit{(j=11/2, l=5)} \\
                                 &   39.78    \textit{(j=13/2, l=6)} \\
\hline
\end{tabular}
\end{center}
\label{table:res}
\end{table}
Those resonances will result in fluctuations of the imaginary part of $V^{RPA}$ and $V^{(2)}$ (Eqs.~\eqref{eq:vrpa} and \eqref{eq:vph}) 
whenever the energy $E-E_{N}$ matches a resonance energy $\varepsilon_{\lambda}$ of the intermediate single-particle state. As a consequence, 
those resonances will strongly influence the corresponding cross section. 
\begin{figure}[h!]
\centering
\adjustbox{trim={0.\width} {0.\height} {0.\width} {0.\height},clip}
{\includegraphics[width=0.45\textwidth,angle=-00,clip=false]{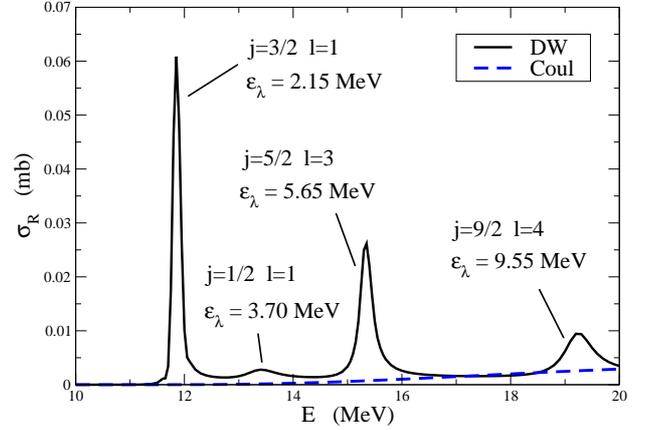}}
\caption{Reaction cross section \textit{vs.} proton incident energy with $V = V^{HF}+\textrm{Im}[V^{RPA}/2]$ coupling only to 
the first 1$^{-}$ excited state in $^{40}$Ca ($E_{x}=9.7$~MeV). Comparison between $\phi_{\lambda}$ treated as a Coulomb 
wave (dashed line) or as a single-particle state in the HF field (solid line).}
\label{fig:dweffect}
\end{figure}
As an illustration of resonance effects, we show, in Fig.~\ref{fig:dweffect}, the reaction cross section for proton scattering 
with a potential including the HF potential as real part and the imaginary part of the RPA potential generated restricting couplings 
to the second $J^{\pi}=1^{-}$ excited state of $^{40}$Ca with $E_{x}=9.7$~MeV excitation energy. This is done taking $\Gamma(E_{N})=0$~MeV 
in Eq.~\eqref{eq:vrpa} in order to emphasize the effect of resonances. This result is compared with a calculation using a Coulomb wave as 
intermediate state. The exact treatment of the intermediate state leads to a global enhancement of the reaction cross section.
\begin{figure}[h!]
\centering
\adjustbox{trim={0.\width} {0.\height} {0.\width} {0.\height},clip}
{\includegraphics[width=0.45\textwidth,angle=-00,clip=false]{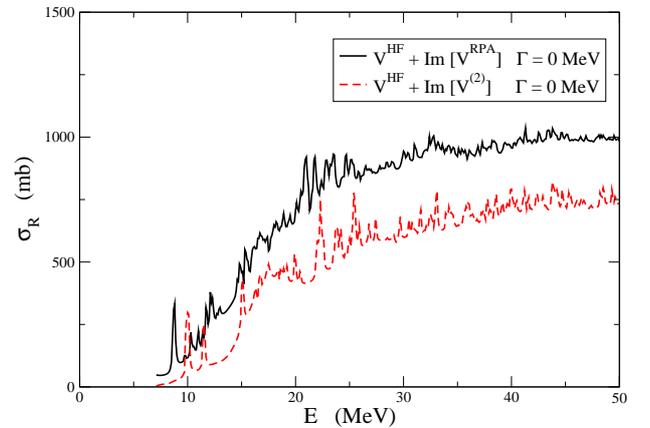}}
\caption{Reaction cross section \textit{vs.} neutron incident energy. Neutron scattering from $^{40}$Ca with $V^{HF}~+~\textrm{Im}~[V^{RPA}]$ 
(solid line) and $V^{HF}~+~\textrm{Im}~[V^{(2)}]$ (dashed line) coupling to all available open channels.}
\label{fig:cross-rpa-width0}
\end{figure}
Moreover we notice that coupling to only one excited state of the target already leads to four resonances between 10 and 20 MeV. One 
expects to get a large number of resonant contributions once coupling to the thousand target excited states. As an example, we show in 
Fig.~\ref{fig:cross-rpa-width0} the same calculation as in Fig.~\ref{fig:dweffect} but including all the open channels for a given 
incident energy. Once again as discussed in Sec.~\ref{sec:formalism}, one needs to average fluctuating contributions before comparing 
scattering observables with experiment. This shows the importance of a complete treatment of the intermediate wave. 

\section{Microscopic potential and cross section data}
\label{sec:cross}
As a first application, NSM has been applied to neutron and proton scattering from $^{40}$Ca using Gogny D1S interaction. 
\begin{figure}[h!]
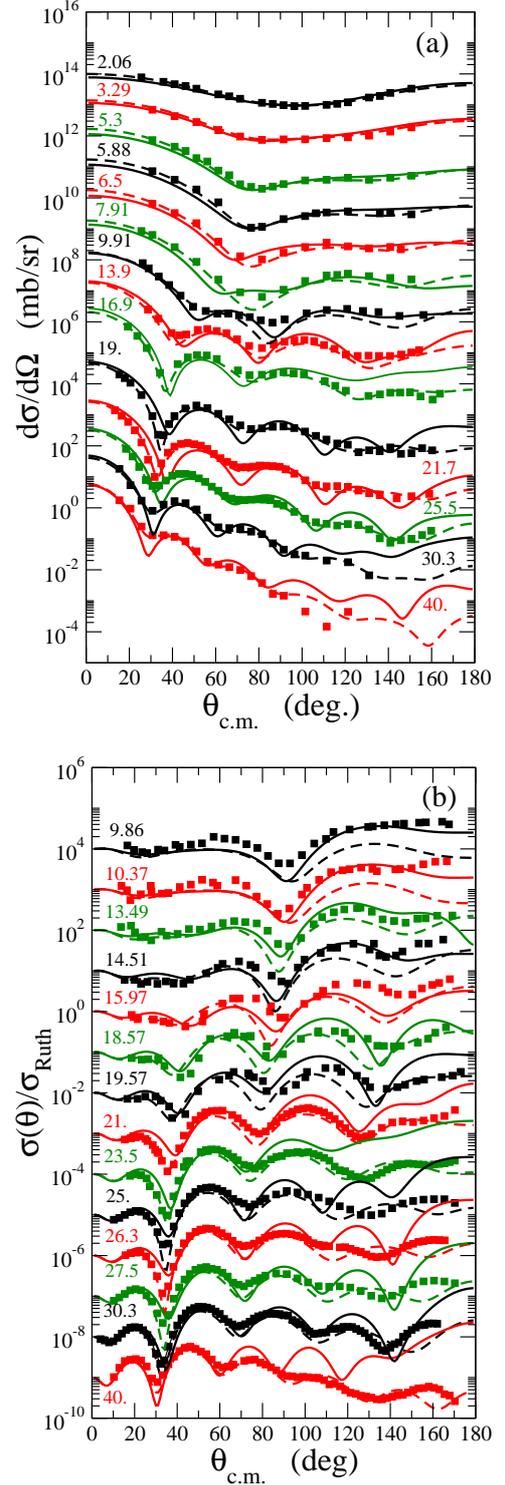

\begin{center}
\adjustbox{trim={0.\width} {0.\height} {0.\width} {0.\height},clip}
{\includegraphics[width=.35\textwidth,angle=-00,clip=false]{fig9a.eps}}
\end{center}
\begin{center}
\adjustbox{trim={0.\width} {0.\height} {0.\width} {0.\height},clip}
{\includegraphics[width=.35\textwidth,angle=-00,clip=right]{fig9b.eps}}
\end{center}
\caption{Differential cross sections for neutron (a) and proton (b) scattering from $^{40}$Ca. Comparison between data 
(symbols), $V^{HF}+\Delta V$ results (solid curves) and Koning-Delaroche potential results (dashed curves).}
\label{fig:sec-diff-rpaall}
\end{figure}
The corresponding differential cross sections for incident energies below 40~MeV are presented in Fig.~\ref{fig:sec-diff-rpaall}. 
Compound-elastic corrections furnished by the Hauser-Feshbach formalism using Koning-Delaroche potential with \textsc{TALYS} are applied to
cross sections obtained from NSM and Koning-Delaroche potential, respectively. It is mostly relevant below 10~MeV for neutron projectile while 
it gives a smaller contribution for proton. NSM results compare very well to experiment and those based on Koning-Delaroche potential up to about 30~MeV 
incident energy. References to data are given in Ref.~\cite{koning_03}. Error bars are smaller than the size symbols. Beyond 
30~MeV, backward-angle cross sections are overestimated. Discrepancies at 16.9~MeV (23.5~MeV) for neutron (proton) scattering are related to 
resonances in the intermediate single-particle state when not completely averaged. A detailed treatment of the width might 
cure this issue.
\begin{figure}[h]
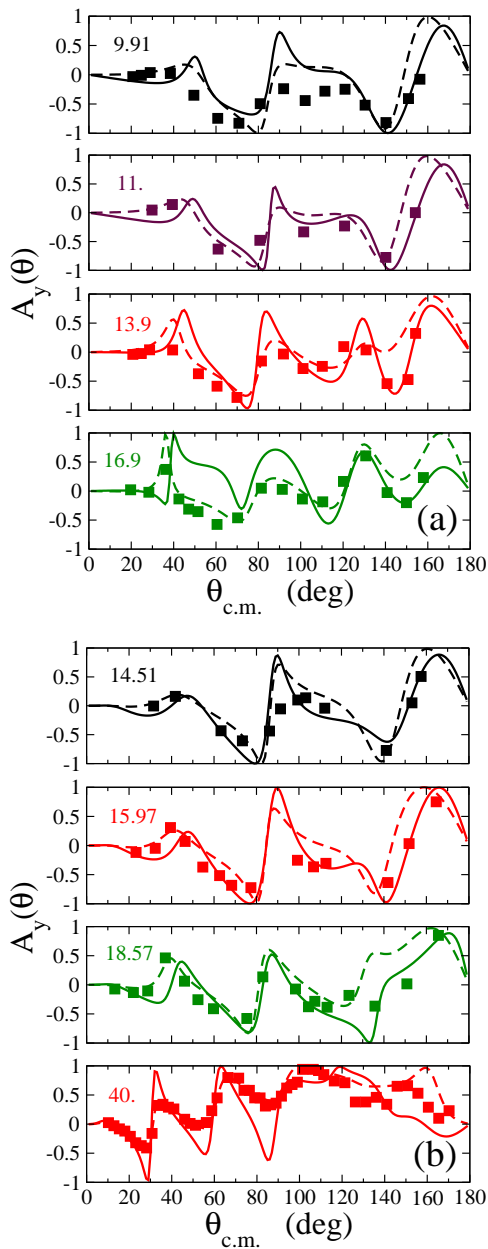

\begin{center}
\adjustbox{trim={0.\width} {0.\height} {0.\width} {0.\height},clip}
{\includegraphics[width=.35\textwidth,angle=-00,clip=false]{fig10a.eps}}
\end{center}
\begin{center}
\adjustbox{trim={0.\width} {0.\height} {0.\width} {0.\height},clip}
{\includegraphics[width=.35\textwidth,angle=-00,clip=false]{fig10b.eps}}
\end{center}
\caption{Same as Fig.~\ref{fig:sec-diff-rpaall} for analyzing powers.}
\label{fig:pol-diff-rpaall}
\end{figure}
In Fig.~\ref{fig:pol-diff-rpaall}, we show calculated analyzing powers for neutron and proton scattering at 
several energies, in good agreement with measurements. Moreover, agreement with data is comparable to that obtained from 
Koning-Delaroche potential. These results suggest that NSM potential retains the correct spin-orbit behavior. 
\begin{figure}[h]
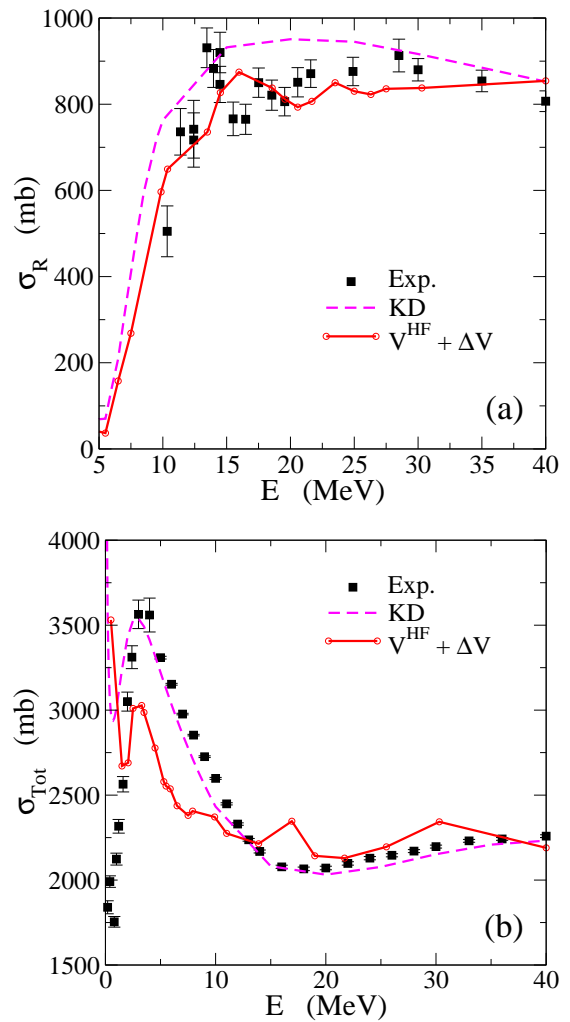

\begin{center}
\adjustbox{trim={0.\width} {0.\height} {0.\width} {0.\height},clip}
{\includegraphics[width=.4\textwidth,angle=-00,clip=false]{fig11a.eps}}
\end{center}
\begin{center}
\adjustbox{trim={0.\width} {0.\height} {0.\width} {0.\height},clip}
{\includegraphics[width=.4\textwidth,angle=-00,clip=false]{fig11b.eps}}
\end{center}
\caption{Reaction cross section for proton (a) and total cross section for neutron (b) scattering from $^{40}$Ca. 
Comparison between data (symbols), $V^{HF}+\Delta V$ results (solid curve) and Koning-Delaroche potential (dashed curve).}
\label{fig:cross-reac}
\end{figure}
In Fig.~\ref{fig:cross-reac}, we show reaction cross section for proton scattering and total cross section for neutron 
scattering. Calculated reaction cross sections are in good agreement with experiments. For neutrons, however, we underestimate 
the total cross section below 10~MeV. Considering that the differential elastic cross sections are well reproduced (see Fig.~\label{fig:sec-diff-rpaall}a), 
this underestimation suggests that part of absorption mechanisms is not accounted for, such as target-excited states beyond RPA, 
double-charge exchange process or an intermediate deuteron formation. 

\noindent This microscopic potential makes the bridge between cross section data and effective interaction. It is worth mentioning 
that we already get nice agreement with data without any adjustable parameter and using an effective interaction, the Gogny D1S 
interaction, originally tailored for structure purposes. This framework opens the way for new effective interactions based both on 
structure and reaction constraints.


\section{Phenomenological potential and microscopic potential}
\label{sec:pot}
\begin{figure}[h!]
\begin{center}
\vspace{-1.cm}
\adjustbox{trim={0.\width} {0.\height} {0.\width} {0.\height},clip}
{\includegraphics[width=.45\textwidth,angle=-90,clip=false]{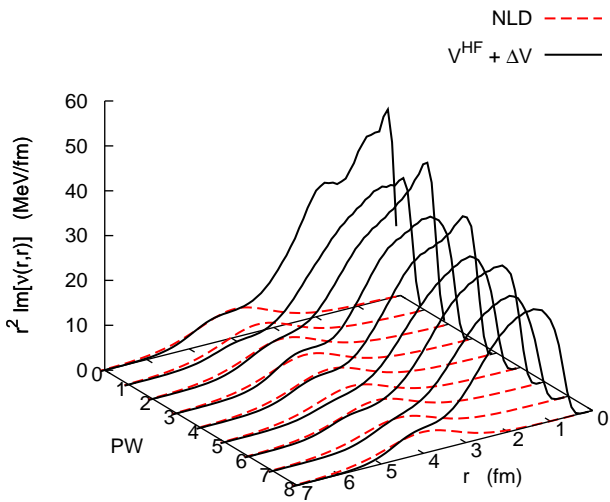}}
\end{center}
\vspace{-1.cm}
\caption{Multipole expansion of the imaginary part of $V^{HF}~+~\Delta V$ (solid curve) and NLD potential (dashed curve) for $r~=~r'$, as 
a function of radius and partial waves for n+$^{40}$Ca at 9.91 MeV.}
\label{fig:pw-40ca-n}
\end{figure}

\begin{figure}[h!]
\hspace{-.7cm}
\adjustbox{trim={0.\width} {0.\height} {0.\width} {0.\height},clip}
{\includegraphics[width=.38\textwidth,angle=-90,clip=false]{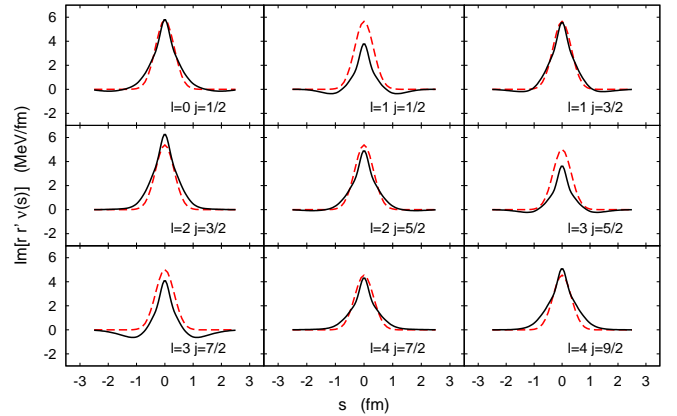}}
\vspace{-.8cm}
\caption{Same as Fig.~\ref{fig:pw-40ca-n} for nonlocality at $r=4.3$~fm.}
\label{fig:pw-40ca-n-nl}
\end{figure}

We shall now highlight the possible crosstalk between microscopic and phenomenological potentials depicted in Fig.~\ref{fig:connect}. 
A large variety of local potentials have been developed in order to describe reaction data. Mahaux and Sartor have then demonstrated the need 
for the potential to satisfy a dispersion relation, connecting its imaginary part to its real part, which provides a link with shell model \cite{mahaux_91}. 
Along that line, local dispersive potentials have been developed \cite{morillon_04,morillon_07}. One issue in those local approaches is the spurious energy dependence 
of the potential coming from the use of a local ersatz to represent a nonlocal object. This issue has been overcome building a dispersive potential with 
a nonlocal static real component \cite{dickhoff_10}. A recent version of this dispersive potential is fully nonlocal both in its real and 
imaginary parts \cite{mahzoon_14}. It is parametrized only for $^{40}$Ca but using all the structure and reaction data available for this nucleus. 
Another nonlocal dispersive potential is currently being developed for a broader range of nuclei \cite{morillon_priv}. 
It is interesting to compare such phenomenological potentials with microscopic and \textit{ab-initio} ones \cite{waldecker_11,dussan_11}. This connection can help 
identifying missing components in microscopic potentials. Reciprocally microscopic potential can provide some guidance for next-to-come potential parametrizations 
regarding for example the shape and the range of the nonlocality or the incident energy dependence. \\
As an illustration, we now consider the case of neutron scattering from $^{40}$Ca at $E=9.91$~MeV. In Fig.~\ref{fig:pw-40ca-n}, we compare the NSM 
potential with the nonlocal dispersive (NLD) potential from Ref.~\cite{mahzoon_14}. We focus on the multipole expansion (see Eq.~\eqref{eq:pw_exp}) 
of the imaginary part of both nonlocal potentials and depict their diagonal contributions. NSM and NLD potentials compare very well around $r=4.3$~fm at 
the nucleus surface. In the volume region, NSM provides a much stronger contribution to the imaginary potential than NLD potential. Nevertheless, NSM leads 
to a reasonable agreement with cross section data as shown in Fig.~\ref{fig:sec-diff-rpaall}. At relatively low incident energy the projectile nucleon is 
only sensitive to the surface region. In other words, low-energy elastic observables do not constrain the volume contribution of the potential. Further inelastic 
scattering calculations using both potentials may help disentangling this discrepancy as partial-wave functions play a major role in such calculations.   
In Fig.\ref{fig:pw-40ca-n-nl}, we compare the nonlocality of both imaginary components at the surface of the nucleus ($r=4.3$~fm) as a function of the nonlocal 
parameter $s=|\textbf{r}-\textbf{r'}|$. We get a good agreement between the microscopic approach and the phenomenological one. Even though small emissive contributions 
appear in some of the multipoles, NSM validates the choice of a Gaussian nonlocality as originally proposed by Perey and Buck \cite{perey_62} and used as 
well for NLD potential. NSM also reproduces the range of the nonlocality of the NLD potential which corresponds to a nonlocality parameter $\beta=0.94$~fm at 
the surface of the nucleus.\\
\noindent We have presented here results only for a given projectile at a given incident energy. A more exhaustive and hopefully conclusive study is in progress.  
Along the same line, it will be interesting to look at the nucleon asymmetry dependence of the NSM potential for example going toward neutron-rich Ca isotopes and 
comparing with the dispersive potential obtained by Charity \textit{et al.} for $^{40,42,44,48,60,70}$Ca \cite{charity_07}.

\section{Phenomenological potential and effective NN interaction}
\label{sec:int}
\begin{figure}[h!]
\centering
\adjustbox{trim={0.\width} {0.\height} {0.\width} {0.\height},clip}
{\includegraphics[width=.4\textwidth,angle=-90,clip=false]{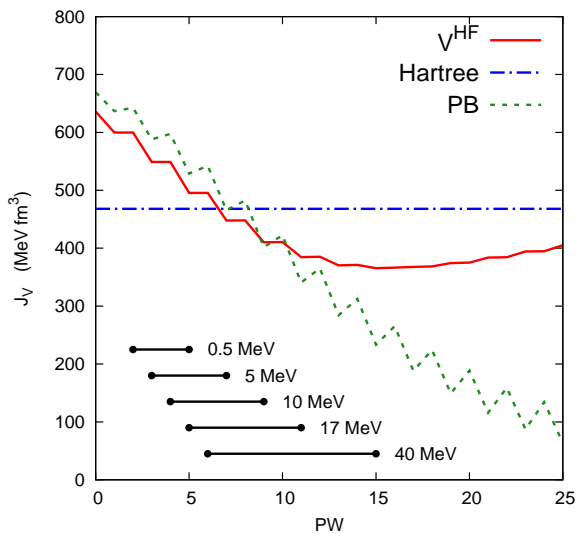}}
\caption{Volume integral as a function of partial waves for neutron scattering from $^{40}$Ca: HF potential (solid curve), Hartree 
potential (dash-dotted curve) and Perey-Buck potential (dotted curve). Horizontal segments denote the partial-wave interval to sum 
up 80\% of the reaction cross section at selected incident energies.}
\label{fig:intvol_n}
\end{figure}

We now consider the connection between phenomenological potential and effective NN interaction through the microscopic potential 
and the structure calculation as shown in Fig.~\ref{fig:connect}. The link between microscopic potential and  phenomenological 
one is done using volume integrals. Volume integrals are useful means of comparison between potentials as they are well constrained 
by scattering data. Here we focus on the real part of the potential. The volume integral for a given multipole $(l,j)$ of the real 
part of the nonlocal potential is defined as,
\begin{eqnarray}
 J^{lj}_{V}(E) &=& \frac{-4\pi}{A} \int dr\ r^{2} \int dr' r'^{2} \textrm{Re} [\nu_{lj}(r,r',E)],
\end{eqnarray}
where $A$ is the nucleon number of the target. The HF potential is the leading contribution to the real part of NSM potential in 
Eq.~\eqref{eq:v}. In Fig.~\ref{fig:intvol_n}, we present the volume integral of the multipole expansion of the HF potential, 
Eq.~\eqref{eq:vhf}, as a function of the partial wave. We compare it to the same quantity obtained from Perey-Buck (PB) nonlocal 
potential \cite{perey_62}. HF potential gives results similar to PB potential up to about the twelfth partial wave. Black segments 
denote the strongest partial-wave contributions accounting for 80\% of the reaction cross section at the selected incident energies. 
Hence taking the PB potential as a reference, the HF potential has a reasonable behavior up to about 17 MeV incident energy. Beyond this energy HF saturates, following 
the trend of the Hartree potential which is local and thus partial-wave independent. As a result increasing incident energy, HF yields 
a much too large volume integral reflected in an overestimate of the differential cross section at backward angles as shown in 
Fig.~\ref{fig:sec-diff-rpaall}. We present here results obtained with the D1S parametrization of the Gogny force, but we came 
to the same conclusions using the D1M parametrization \cite{goriely_09}. 
The behavior of the volume integral as a function of the partial wave is dictated by the shape and the range of the nonlocality. PB potential is built 
with a Gaussian nonlocality whereas the HF potential is made of a local Hartree term and a nonlocal Fock term as shown in Eq.~\eqref{eq:vhf}. 
Those two contributions can be related to the different terms of the effective NN interaction. Gogny interaction is built with two Gaussian ranges 
in its central part. 
\begin{figure}[h*]
\begin{minipage}[c]{0.45\linewidth}
\begin{center}
\adjustbox{trim={0.\width} {0.\height} {0.\width} {0.\height},clip}
{\includegraphics[width=1.\textwidth,angle=-90,clip=false]{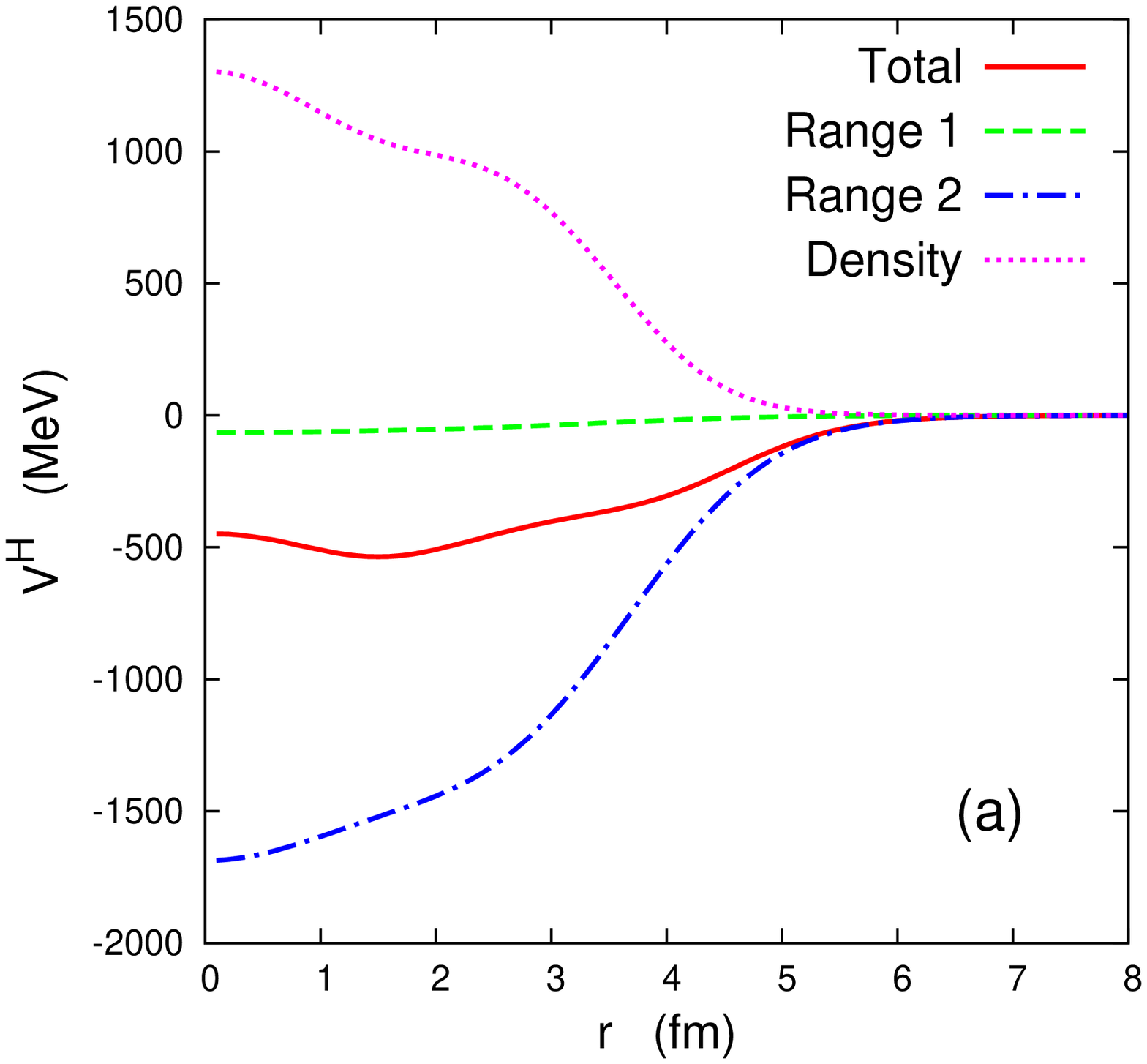}}
\adjustbox{trim={0.\width} {0.\height} {0.\width} {0.\height},clip}
{\includegraphics[width=1.\textwidth,angle=-90,clip=false]{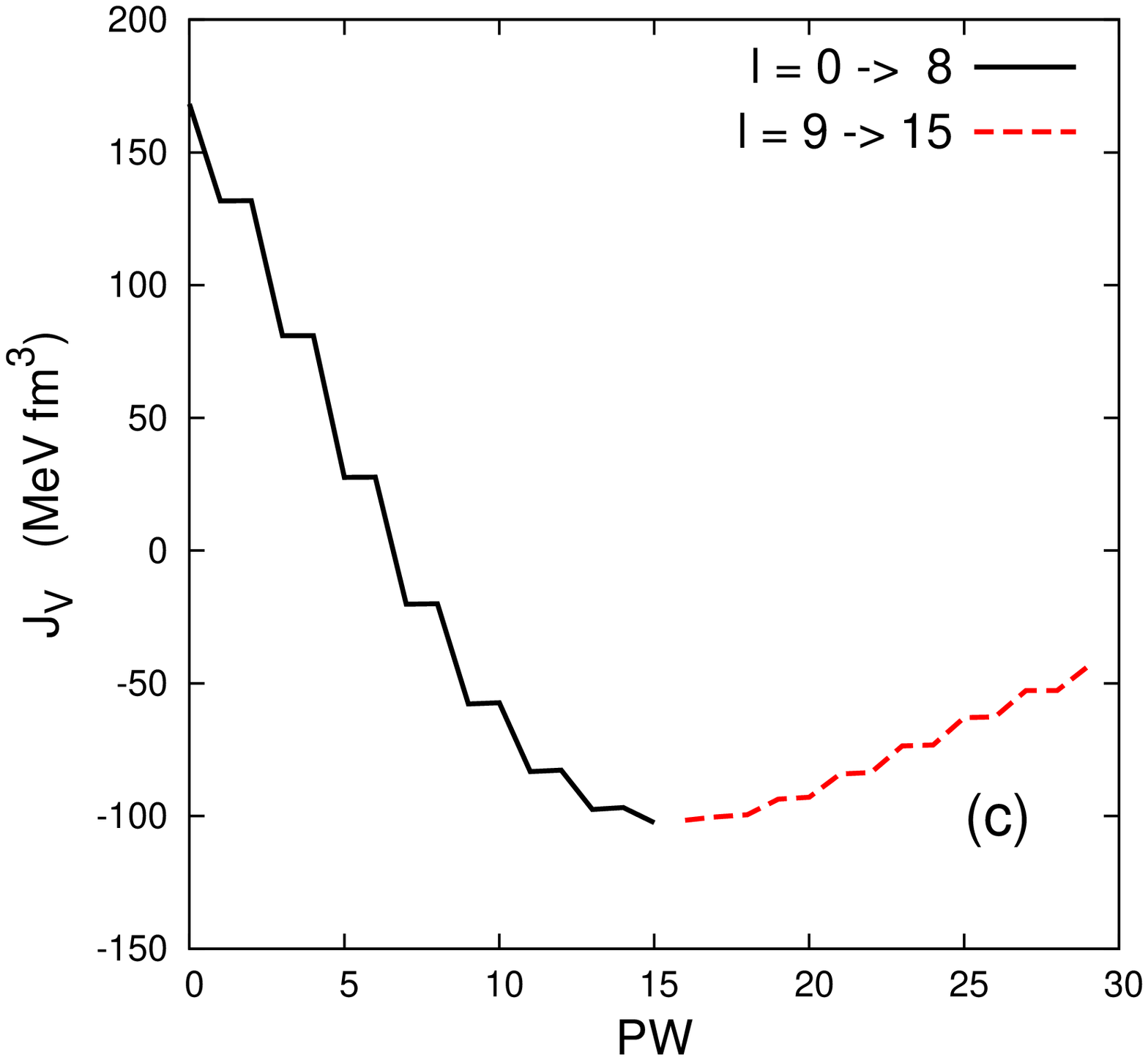}}
\end{center}
\end{minipage}
\hspace{0.3cm}
\begin{minipage}[c]{0.45\linewidth}
\begin{center}
\adjustbox{trim={0.\width} {0.\height} {0.\width} {0.\height},clip}
{\includegraphics[width=1.\textwidth,angle=-90,clip=false]{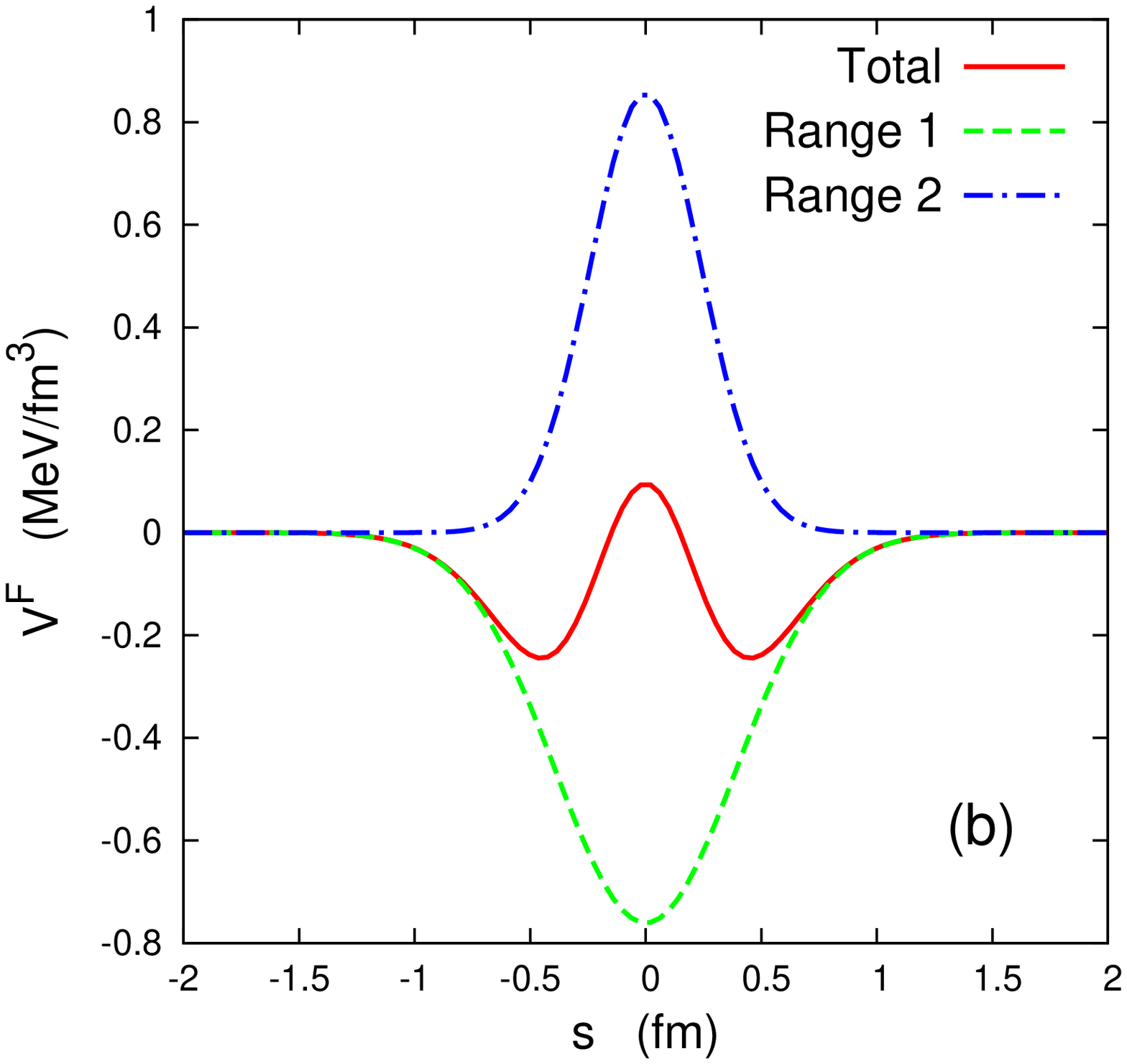}}
\adjustbox{trim={0.\width} {0.\height} {0.\width} {0.\height},clip}
{\includegraphics[width=1.\textwidth,angle=-90,clip=false]{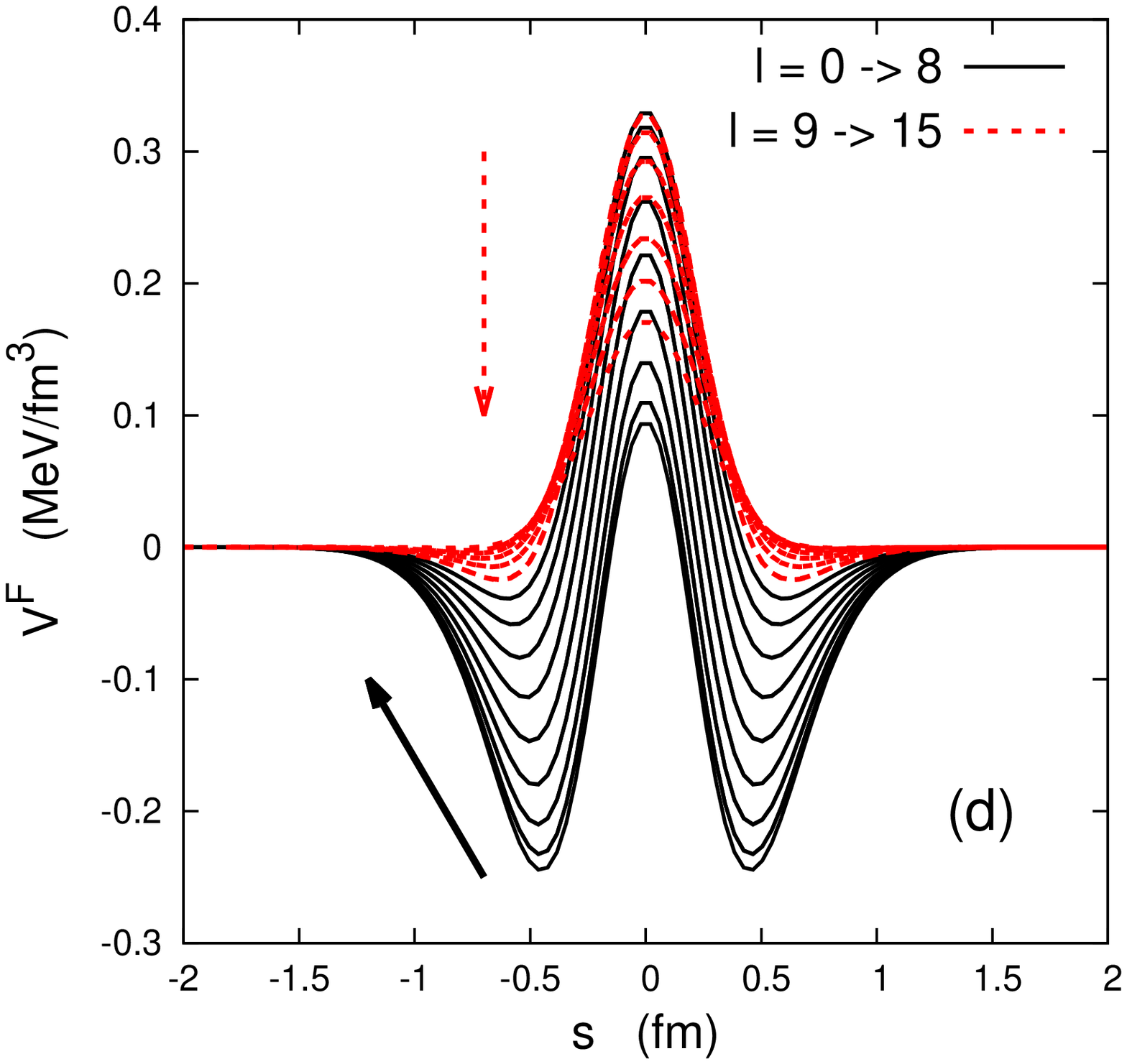}}
\end{center}
\end{minipage}
\caption{Contributions for n + $^{40}$Ca to: (a) to the Hartree local potential ($V^{H}$): Total (solid line), first range of D1S 
(dashed line), second range of D1S (dash-dotted line) and density term (dotted line). (b) First partial wave of the nonlocal 
Fock term at $r=r'=4.3$~fm: Total (solid line), first range of D1S (dashed line) and second range of D1S (dash-dotted line). (c) Volume 
integral of the Fock potential as a function of partial wave: Negative slope (solid line), positive slope (dashed line). (d) Same 
as (c) for the Fock components nonlocality at $r~=~r'~=~4.3$~fm.}
\label{fig:hf}
\end{figure}
The contributions to the Hartree potential of those two Gaussians as well as the contribution of the density-dependent term of the interaction are shown
in panel~(a) of Fig.~\ref{fig:hf}. In panel~(b), we show the contributions of the two central terms with finite range to the nonlocality of 
the first partial wave of the Fock term. The summation of those two Gaussian nonlocalities with opposite sign yields the "W" shape of the total nonlocality. 
We then present in panel~(c) the volume integral corresponding to the nonlocal Fock potential. The good behavior of the HF volume integral for low partial 
waves is related to the downward slope of the Fock volume integral and its change of sign around the seventh partial wave. Then for higher partial 
waves, the Fock volume integral converges to the zero limit. The corresponding behavior of the Fock term as a function of the partial wave is depicted in 
panel~(d).\\
We plan to investigate to what extent the effective interaction could be improved in order to get a better behavior of the HF potential above 30~MeV. This 
issue emerges from the use of a local and energy-independent ersatz for the effective NN interaction instead of the actual {g}-matrix with its full 
complexity. One way to get a fully nonlocal HF potential would be to use a nonlocal effective NN interaction as proposed for example by Tabakin \cite{tabakin_64}.

\section{Next decade}
\label{sec:next}

The present work constitutes a promising step forward aimed to a model keeping at the same footing both reaction and structure aspects of the many-nucleon 
system. Starting from an effective NN interaction, NSM accounts reasonably well for low-energy scattering data. We use consistently the Gogny D1S interaction, 
although this scheme can be applied to any interaction of similar nature. An important feature of the approach is the extraction of the imaginary part of the potential by means 
of intermediate excitations of the target. The study has been restricted to closed-shell target but can be extended to account for pairing correlations as well as 
axial deformation. We now expose in more detail our plans for the forthcoming years. 

\subsection{Spherical target nuclei}
In the short-term future, we wish to investigate the NSM scheme for spherical targets according to the following plan,

\begin{itemize}
 \item NSM will be applied to a broader range of target nuclei well described within RPA including $^{48}$Ca, $^{90}$Zr, $^{132}$Sn and $^{208}$Pb.
 \item The link between NSM potential and phenomenology can be carried on exploring energy, multipole and mass dependences of the 
 potential. The NSM potential can also provide some trends for the shape and the range of the nonlocality. We plan to investigate 
 the building of surface and volume contributions as a function of target excitations. 
 \item Above about 50~MeV incident energy, a connection can be established between the NSM potential and the folding potential relying on {g}-matrix and 
 thus with the bare NN interaction. This can be a fertile ground for new effective interactions. 
 \item At low incident energy, NSM provides a volume part of the imaginary potential larger than phenomenology. The small volume contribution in phenomenological 
 potential is often justified by the fact that the projectile does not have sufficient energy to knock out a target nucleon. This discrepancy is possibly due to 
 the fact that at low energy the volume part of the imaginary potential is not well constrained because the projectile does not explore the interior of the potential. 
 We plan to use the NSM potential in inelastic scattering calculations in order to disentangle this issue.
 \item In its present version, NSM requires a phenomenological width. This width has several microscopic origins, as discussed in Sec.~\ref{sec:formalism}. A 
 microscopic account of those widths is planed using continuum RPA \cite{dedonno_11} and multihole-multiparticle configuration mixing \cite{pillet_08} for the escape 
 and the damping widths, respectively. 
 \item The NSM potential will be used to provide transmission coefficients for compound-elastic calculations. Moreover, we plan to develop a compound-nucleus 
 formalism based only on NSM. Indeed, NSM gives access to the fluctuating contribution of the S-matrix and as a consequence to the compound-elastic contribution. 
 \item The study of the volume integral of the real part of the potential has exhibited the possible crosstalk between phenomenological potentials and 
 effective NN interactions. In particular, the interaction is not well suited for the description of partial waves with more than about $\ell=7$. 
 Those reaction constraints will be used for new parametrizations of the interaction. Moreover, a nonlocal version of the effective NN interaction could tackle 
 the issue of the saturation of the HF volume integral to the Hartree one. 
\end{itemize}

\subsection{Spherical target nuclei with pairing correlations}
The main next step will be to take into account pairing correlations in spherical nuclei. 

\begin{itemize}
 \item We plan to develop a HFB potential in coordinate space. The mean-field and the pairing field have already been studied in coordinate space in a previous work on 
 Cooper's pairs \cite{pillet_07}. The goal is then to deal with quasiparticle scattering with a special care of resonances in both mean-field and pairing channel as shown 
 in Ref. \cite{bennaceur_99,grasso_01}. 
 \item In the present approach, the intermediate particle has the same nucleonic nature than the incident and the outgoing particle. 
 Previous studies by Osterfeld \textit{et al.} \cite{osterfeld_81} have shown the importance of double charge-exchange.  
 This process can be accounted for in a consistent way using HFB \cite{martini_14a}.
 \item The target excited states will then be described within QRPA \cite{peru_08}. 
\end{itemize}

\subsection{Deformed target nuclei}
In the midterm future, we plan to deal with axially-deformed targets according to the following plan, 

\begin{itemize}
 \item This will require the development of an axially-deformed HFB potential in coordinate space. The corresponding mean-field and the pairing field have 
 already been studied in coordinate space in a previous work on Cooper's pairs \cite{pillet_10}.
 \item QRPA \cite{peru_08} will be used to generate excited states in the intrinsic frame of the target.
 \item A projection an good angular momentum, using the rotational approximation \cite{bohr_98}, will provide the monopole and different coupling potentials 
  to model nucleon elastic scattering from axially deformed target. 
 \item The problem of solving coupled equations with nonlocal potentials will have to be addressed.  
\end{itemize}

\section*{Acknowledgments}
\noindent H. F. A. acknowledges partial funding from FONDECYT under Grant No 1120396.

\bibliographystyle{epj}

\end{document}